\documentclass[%
reprint,
superscriptaddress,
showpacs,
preprintnumbers,
amsmath,
amssymb,
aps,
prd,
]{revtex4-1}

\usepackage{graphicx}
\usepackage{dcolumn}
\usepackage{bm}
\def\a{\alpha}
\def\b{\beta}

\def\d{\delta}
\def\e{\epsilon}

\def\g{\gamma}
\def\p{\psi}

\def\r{\rho}
\def\s{\sigma}

\def\be{\begin{equation}}
\def\ee{\end{equation}}
\def\arr{\begin{array}{rll}}
\def\ea{\end{array}}
\def\bea{\begin{eqnarray}}
\def\eea{\end{eqnarray}}

\begin{document}

\title{$\mathcal{N}=4$ super--Schwarzian via nonlinear realizations}
\author{Anton Galajinsky}
\email{a.galajinsky@tusur.ru}
\affiliation{Tomsk State University of Control Systems and Radioelectronics, 634050 Tomsk, Russia}
\author{Sergey Krivonos}
\email{krivonos@theor.jinr.ru}
\affiliation{Bogoliubov  Laboratory of Theoretical Physics, JINR,
141980 Dubna, Russia}

\begin{abstract}
Current studies of supersymmetric extensions of the Sachdev--Ye--Kitaev model stimulate a renewed interest in super--Schwarzian derivatives. In this work, we
apply the method of nonlinear realizations to the finite--dimensional superconformal group $SU(1,1|2)$ and link its invariants to the $\mathcal{N}=4$ super--Schwarzian.
\end{abstract}

\maketitle

\section{\label{sec:I} Introduction}

Current studies of supersymmetric extensions of the Sachdev--Ye--Kitaev model (see \cite{FGMS,MTV,BBN} and references therein)
stimulate a renewed interest in super--Schwarzian derivatives \cite{F,Cohn,Sch,MU}. An $\mathcal{N}$--extended super--Schwarzian acts upon a fermionic superfield which specifies superconformal diffeomorphisms of the odd sector of $\mathcal{S}^{1|\mathcal{N}}$ superspace \cite{footnote}.
It enjoys a remarkable composition law which implies invariance of the $\mathcal{N}=1,2,3,4$ super--Schwarzian under finite transformations forming $OSp(1|2)$, $SU(1,1|1)$, $OSp(3|2)$, and $SU(1,1|2)$ superconformal group, respectively.

A conventional way of introducing a super--Schwarzian derivative is to compute a (finite) superconformal transformation of the super stress--energy tensor underlying a $2d$ $\mathcal{N}$--extended conformal field theory, in which it shows up as the anomalous term \cite{F,Cohn,Sch,MU}. Alternatively, one can study the cocycles describing central extensions of infinite dimensional Lie superalgebras (see, e.g., \cite{MD}). Because for $\mathcal{N}\geq 5$ the construction of the central term operator is problematic \cite{ChK}, the $\mathcal{N}=1,2,3,4$ instances mentioned above seem to exhaust all available options.

In a recent work \cite{AG}, a third alternative was studied, which consists in applying the method of nonlinear realizations \cite{CWZ} to finite--dimensional superconformal groups. In this setting,
a super--Schwarzian derivative is linked to the supergroup invariants.

Within the method of nonlinear realizations, one usually starts with a coset space element $\tilde g$, on which a (super)group representative $g$ acts by the left multiplication $\tilde g'=g \cdot \tilde g$. Then one constructs the Maurer--Cartan one--forms
${\tilde g}^{-1} d \tilde g$, where $d$ is the (super)differential, which automatically hold invariant under the transformation. These invariants can be used to impose constraints allowing one to express some of  (super)fields parametrizing the coset element $\tilde g$ in terms of the other \cite{IO}. If the algebra at hand is such that all but one (super)fields can be linked to a single unconstrained (super)field, then the last remaining Maurer--Cartan invariant describes a derivative of the latter, which holds invariant under the action of the (super)group one started with. In particular, in \cite{AG} the $\mathcal{N}=1$ and $\mathcal{N}=2$ super--Schwarzian derivatives were obtained by applying the procedure to $OSp(1|2)$ and $SU(1,1|1)$ superconformal groups, respectively. A non--sypersymmetric case was previously studied in \cite{AG1}.

The goal of this work is to provide a similar derivation of the $\mathcal{N}=4$ super--Schwarzian associated with $SU(1,1|2)$ superconformal group.

The paper is organized as follows. In the next section, superconformal diffeomorphisms of $\mathcal{S}^{1|4}$ superspace are considered and conditions which follow from the requirement that the covariant derivatives transform homogeneously are analysed. For the fermionic superfield, which describes superconformal diffeomorphisms of the odd sector of $\mathcal{S}^{1|4}$, one reveals the chirality condition and an extra quadratic constraint \cite{Sch}. The restrictions are solved explicitly and it is demonstrated that the latter is actually equivalent to a simpler linear equation (see Eq. (\ref{Const6}) below).

In Sect. ~\ref{sec:III}, the method of nonlinear realizations is applied to the finite--dimensional superconformal group $SU(1,1|2)$ with the aim to link its invariants to an $\mathcal{N}=4$ super--Schwarzian. First, each generator in $su(1,1|2)$ superalgebra is accompanied by a Goldstone superfield of the same Grassmann parity, which all together give rise to a group--theoretic element $\tilde g$. Then the Maurer--Cartan invariants ${\tilde g}^{-1} \mathcal{D}^\alpha \tilde g$, where $\mathcal{D}^\alpha $ is the covariant derivative, are computed. After that, constraints are imposed, which enable one to link all the Goldstone superfields entering $\tilde g$ to a single fermionic superfield (a companion of the supersymmetry generator). Substituting the resulting relations back into the Maurer--Cartan invariants, one unambiguously reproduces the $\mathcal{N}=4$ super--Schwarzian derivative introduced in \cite{MU}. Finally, properties of the super--Schwarzian, including the finite $SU(1,1|2)$ transformations which leave it invariant, are discussed.
In the concluding Sect.~\ref{sec:IV} we summarise our results and discuss possible further developments. Our spinor conventions are gathered in Appendix.

Throughout the text summation over repeated indices is understood.

\section{\label{sec:II}   Superconformal diffeomorphisms of $\mathcal{S}^{1|4}$}

Consider $\mathcal{S}^{1|4}$ superspace parametrized by a real bosonic coordinate $t$ and a pair of Hermitian conjugate
anti--commuting $SU(2)$--spinors $(\theta_\alpha,\bar\theta^\alpha)$, ${(\theta_\alpha)}^{\dagger}=\bar\theta^\alpha$, $\alpha=1,2$ (see Appendix for our spinor conventions). The $d=1$, $\mathcal{N}=4$ supersymmetry algebra
\be
\{q_\alpha, {\bar q}^\beta\}=2 h {\delta_\alpha}^\beta
\ee
allows one to represent $\mathcal{S}^{1|4}$ as the supergroup manifold
\be
\tilde g=e^{i t h} e^{\theta^\alpha q_\alpha+{\bar\theta}_\alpha {\bar q}^\alpha}.
\ee
The left action of the supergroup on the superspace, $\tilde g'=e^{i a h} e^{\epsilon^\alpha q_\alpha+{\bar\epsilon}_\alpha {\bar q}^\alpha} \cdot \tilde g$, where $a$ and $(\epsilon^\alpha,{\bar\epsilon}_\alpha)$ are even and odd supernumbers, respectively, generates the $d=1$, $\mathcal{N}=4$ supersymmetry transformations
\bea
&&
t'=t+a; \qquad {\theta}'_\alpha=\theta_\alpha+\epsilon_\alpha, \qquad  {\bar\theta}'^\alpha={\bar\theta}^\alpha+{\bar\epsilon}^\alpha,
\nonumber\\[2pt]
&&
t'=t-i\left(\epsilon_\alpha {\bar\theta}^\alpha+{\bar\epsilon}^\alpha \theta_\alpha \right).
\eea

Covariant derivatives, which anticommute with the supersymmetry generators, read
\be
{\mathcal{D}}^\alpha=\partial^\alpha+i {\bar\theta}^\alpha \partial_t, \qquad {\bar{\mathcal{D}}}_\alpha={\bar\partial}_\alpha+i \theta_\alpha \partial_t,
\label{covd}
\ee
where $\partial_t=\frac{\partial}{\partial t}$, $\partial^\alpha=\frac{\vec{\partial}}{\partial \theta_\alpha}$, ${\bar\partial}_\alpha=\frac{\vec{\partial}}{\partial \bar\theta^\alpha}$. They satisfy the relations
\bea
&&
\{ {\mathcal{D}}^\alpha, {\bar{\mathcal{D}}}_\beta \}=
2i {\delta_\beta}^\alpha \partial_t, \qquad
{\mathcal{D}}^\alpha {\mathcal{D}}^\beta=-\frac 12 \epsilon^{\alpha\beta} {\mathcal{D}}^2,
\nonumber\\[2pt]
&&
 {\bar{\mathcal{D}}}_\alpha {\bar{\mathcal{D}}}_\beta=-\frac 12 \epsilon_{\alpha\beta} {\bar{\mathcal{D}}}^2, \qquad
[{\mathcal{D}}^2,{\bar{\mathcal{D}}}_\alpha]=-4i {\mathcal{D}}_\alpha \partial_t,
\nonumber\\[2pt]
&&
[{\bar{\mathcal{D}}}^2,{\mathcal{D}}^\alpha]=-4i {\bar{\mathcal{D}}}^\alpha \partial_t,
\nonumber\\[2pt]
&&
[{\bar{\mathcal{D}}}^2,{\mathcal{D}}^2]=-4i\left({\mathcal{D}}^\alpha {\bar{\mathcal{D}}}_\alpha-{\bar{\mathcal{D}}}_\alpha {\mathcal{D}}^\alpha \right) \partial_t,
\label{pr}
\eea
with ${\mathcal{D}}^2={\mathcal{D}}^\alpha {\mathcal{D}}_\alpha$, ${\bar{\mathcal{D}}}^2={\bar{\mathcal{D}}}_\alpha {\bar{\mathcal{D}}}^\alpha$.

In what follows, we will need a component decomposition of a chiral fermionic superfield $\psi_\beta$, which obeys the equation
\be
{\bar{\mathcal{D}}}_\beta \psi_\gamma=0.
\label{chc}
\ee
Taking into account the identity ${\bar{\mathcal{D}}}_\beta=e^{-i\bar\theta \theta \partial_t} \bar\partial_\beta e^{i\bar\theta \theta \partial_t}$,
one gets
\bea
&&
\psi_\gamma(t,\theta,\bar\theta)=\alpha_\gamma(t)+\theta_\beta {b_\gamma}^\beta (t)-i\bar\theta\theta {\dot\alpha}_\gamma (t)+\theta^2 \beta_\gamma (t)+
\nonumber\\[2pt]
&&
\qquad \qquad \qquad
\frac{i}{2}\bar\theta_\beta \theta^2 {{\dot b}_\gamma}~^\beta (t)-\frac 14 \theta^2 \bar\theta^2 {\ddot\alpha}_\gamma (t),
\label{csf}
\eea
where $\alpha_\gamma (t)$, $\beta_\gamma (t)$ are complex fermionic components and ${{b}_\gamma}^\beta (t)$ is a complex bosonic matrix--valued function of $t$.
The Hermitian
conjugation rules
\bea
&&
 {\left(\mathcal{D}^\alpha\rho\right)}^{\dagger}=-\bar{\mathcal{D}}_\alpha \rho, \qquad {\left(\mathcal{D}^\alpha \psi_\beta\right)}^{\dagger}=\bar{\mathcal{D}}_\alpha \bar\psi^\beta,
\nonumber\\[2pt]
&&
{\left(\mathcal{D}^\alpha \psi^\beta\right)}^{\dagger}=-\bar{\mathcal{D}}_\alpha \bar\psi_\beta,
\eea
which involve a real bosonic superfield $\rho$, a complex fermionic superfield $\psi_\alpha$, and its Hermitian conjugate $\bar\psi^\alpha={(\psi_\alpha)}^{\dagger}$, will be heavily used below.

Similarly to the $\mathcal{N}=1$ and $\mathcal{N}=2$ cases \cite{F,Cohn} (see also \cite{IKL1}), superconformal diffeomorphisms of $\mathcal{S}^{1|4}$ are introduced as the transformations
\be
t'=\rho(t,\theta,\bar\theta), \quad \theta'_\alpha=\psi_\alpha (t,\theta,\bar\theta), \quad \bar\theta'^\alpha=\bar\psi^\alpha (t,\theta,\bar\theta),
\label{sdiff}
\ee
where $\rho$ is a real bosonic superfield and $\psi_\alpha$ is a complex fermionic superfield,
under which the covariant derivatives transform homogeneously
\be
{\mathcal{D}}^\alpha=\left( {\mathcal{D}}^\alpha \psi_\beta\right) {\mathcal{D}}'^\beta, \qquad \bar{\mathcal{D}}_\alpha=\left(\bar{\mathcal{D}}_\alpha \bar\psi^\beta \right) \bar{\mathcal{D}}'_\beta.
\label{hom1}
\ee
Eq. (\ref{hom1}) yields the constraints
\bea
&&
\bar{\mathcal{D}}_\alpha \psi_\beta=0, \qquad {\mathcal{D}}^\alpha \rho-i \left({\mathcal{D}}^\alpha \psi_\beta\right)\bar\psi^\beta=0,
\nonumber\\[2pt]
&&
{\mathcal{D}}^\alpha \bar\psi^\beta=0, \qquad \bar{\mathcal{D}}_\alpha \rho-i\left(\bar{\mathcal{D}}_\alpha \bar\psi^\beta \right) \psi_\beta=0,
\label{Const1}
\eea
which also imply
\be
\partial_t \rho=-i \bar\psi^\alpha \partial_t \psi_\alpha+i \partial_t \bar\psi^\alpha \psi_\alpha+\frac 12  \left(\mathcal{D} \psi \bar{\mathcal{D}} \bar\psi\right),
\label{cond1}
\ee
where we abbreviated $\mathcal{D}^\alpha \psi_\beta \bar{\mathcal{D}}_\alpha \bar\psi^\beta=\left(\mathcal{D} \psi \bar{\mathcal{D}} \bar\psi\right)$. Thus, $\rho$ is fixed provided $\psi_\alpha$ is known.

The compatibility of (\ref{hom1}) with the properties of the covariant derivatives (\ref{pr}) imposes further restrictions on $\psi_\alpha$. From $\{ {\mathcal{D}}^\alpha, {\bar{\mathcal{D}}}_\beta \}=
2i {\delta_\beta}^\alpha \partial_t$ and the identity
\be
\partial_t=\partial_t \psi_\alpha  {\mathcal{D}'}^\alpha+\partial_t \bar\psi^\alpha \bar{\mathcal{D}'}_\alpha+\frac 12 \left(\mathcal{D} \psi \bar{\mathcal{D}} \bar\psi\right) \partial_{t'},
\ee
one gets the quadratic constraint \cite{Sch}
\be
{\mathcal{D}}^\alpha \psi_\lambda \bar{\mathcal{D}}_\beta \bar\psi^\lambda=\frac 12 {\delta_\beta}^\alpha \left(\mathcal{D} \psi \bar{\mathcal{D}} \bar\psi\right).
\label{Const3}
\ee
Thus, up to a factor, ${\mathcal{D}}^\alpha \psi_\lambda$ is a unitary matrix, which also implies
\bea
&&
{\mathcal{D}}^\lambda \psi_\alpha \bar{\mathcal{D}}_\lambda \bar\psi^\beta=\frac 12 {\delta_\alpha}^\beta \left(\mathcal{D} \psi \bar{\mathcal{D}} \bar\psi\right) \quad \Rightarrow
\nonumber\\[2pt]
&&
{\mathcal{D}}^\alpha \psi_\beta=-\frac 12 \bar{\mathcal{D}}^\alpha \bar\psi_\beta \frac{\left(\mathcal{D} \psi \bar{\mathcal{D}} \bar\psi\right)}{\mbox{det} \left(\bar{\mathcal{D}} \bar\psi\right) },
\label{Const5}
\eea
with $\mbox{det} \left( \bar{\mathcal{D}} \bar\psi\right)=-\frac 12 \epsilon^{\alpha\beta} \epsilon_{\gamma\delta} \bar{\mathcal{D}}_\alpha \bar\psi^\gamma \bar{\mathcal{D}}_\beta \bar\psi^\delta$.
Computing the covariant derivatives of (\ref{Const3}), one gets a chain of relations, two of which
\bea
&&
\mathcal{D}^\alpha \left(\mathcal{D} \psi \bar{\mathcal{D}} \bar\psi \right) =4i \partial_t \bar\psi^\beta \mathcal{D}^\alpha \psi_\beta,
\nonumber\\[2pt]
&&
\bar{\mathcal{D}}_\alpha \left(\mathcal{D} \psi \bar{\mathcal{D}} \bar\psi \right)=4i \partial_t \psi_\beta \bar{\mathcal{D}}_\alpha  \bar\psi^\beta
\label{pr1}
\eea
will be important for what follows. Because the last two relations in (\ref{pr}) result from $\{ {\mathcal{D}}^\alpha, {\bar{\mathcal{D}}}_\beta \}=
2i {\delta_\beta}^\alpha \partial_t$, they do not produce further restrictions on $\psi_\alpha$.

Using the covariant projection method, in which components of a superfield are linked to its covariant derivatives evaluated at $\theta_\alpha=0$, $\bar\theta^\alpha=0$, one can verify that Eq. (\ref{Const3})
connects the fermionic components of the chiral superfield (\ref{csf}) to each other and reduces the matrix--valued bosonic function ${b_\alpha}^\beta (t)$  to a single unknown scalar
\bea
&&
\beta_\gamma(t)=i {\dot{\bar\alpha}}_\gamma (t) e^{2i v},
\nonumber\\[2pt]
&&
{b_\alpha}^\beta (t)=u(t) e^{i v} {{\left(\mbox{exp} \left[\frac{i}{2} \xi_c \sigma_c\right] \right)}_\alpha}^\beta.
\label{funct}
\eea
Here $u(t)$ is an arbitrary real function of $t$, $v$ and $\xi_a$, $a=1,2,3$, are real bosonic constants, and $\sigma_a$ are the Pauli matrices (see Appendix).

Substituting (\ref{funct}) into the covariant derivatives of $\psi_\alpha$ and $\bar\psi^\alpha$, one reveals the identity
\be
e^{-iv}{\mathcal{D}}^\alpha \psi_\beta+e^{iv} \bar{\mathcal{D}}^\alpha \bar\psi_\beta=0.
\ee
Because one can always redefine the fermionic superfield $e^{-iv} \psi_\alpha \to \psi_\alpha$, in what follows we set the parameter $v$ in (\ref{funct}) to vanish, thus reducing the quadratic constraint (\ref{Const3}) to the linear restriction
\be
{\mathcal{D}}^\alpha \psi_\beta+ \bar{\mathcal{D}}^\alpha \bar\psi_\beta=0.
\label{Const6}
\ee
The latter also implies ${\mathcal{D}}^2 \psi_\alpha=-4 i\partial_t \bar\psi_\alpha$. The fact that the quadratic equation (\ref{Const3}) is equivalent to (\ref{Const6}) seems to have escaped attention thus far.

The Taylor series expansion of $u(t)$ and $\alpha_\gamma (t)$ involves an infinite number of constant parameters, which all together represent an infinite--dimensional extension of $SU(1,1|2)$.

\section{\label{sec:III}  $\mathcal{N}=4$ super--Schwarzian via nonlinear realizations}

As was mentioned in the Introduction, the goal of this work is to
link the $\mathcal{N}=4$ super--Schwarzian derivative to invariants of $SU(1,1|2)$ superconformal group.
To this end, let us first consider the structure relations of the superconformal algebra $su(1,1|2)$
\bea
&&
[P,D]=i P, \quad [P,K]=2iD,
\nonumber\\[10pt]
&&
[D,K]=iK, \quad [\mathcal{J}_a,\mathcal{J}_b]=i\epsilon_{abc} \mathcal{J}_c,
\nonumber\\[10pt]
&&
\{ Q_\a, \bar Q^\b \}=2P {\d_\a}^\b, \quad
\{ Q_\a, \bar S^\b \}=2i {{(\s_a)}_\a}^\b \mathcal{J}_a-2D {\d_\a}^\b,
\nonumber\\[10pt]
&&
\{ S_\a, \bar S^\b \}=2K {\d_\a}^\b, \quad
\{ \bar Q^\a, S_\b \}=-2i{{(\s_a)}_\b}^\a \mathcal{J}_a-2D {\d_\b}^\a,
\nonumber\\[6pt]
&&
 [D,Q_\a] = -\frac{i}{2} Q_\a, \quad [D,S_\a] =\frac{i}{2} S_\a,
\nonumber\\[6pt]
&&
[K,Q_\a] =iS_\a, \quad [P,S_\a]=-iQ_\a,
\nonumber\\[6pt]
&&
[\mathcal{J}_a,Q_\a] =-\frac{1}{2} {{(\s_a)}_\a}^\b Q_\b, \quad [\mathcal{J}_a,S_\a] =-\frac{1}{2} {{(\s_a)}_\a}^\b S_\b,
\nonumber\\[6pt]
&&
 [D,\bar Q^\a]=-\frac{i}{2} \bar Q^\a, \quad [D,\bar S^\a] =\frac{i}{2} \bar S^\a,
\nonumber\\[6pt]
&&
[K,\bar Q^\a] =i\bar S^\a, \quad [P,\bar S^\a] =-i\bar Q^\a,
\nonumber\\[6pt]
&&
[\mathcal{J}_a,\bar Q^\a] =\frac{1}{2} \bar Q^\b {{(\s_a)}_\b}^\a, \quad [\mathcal{J}_a,\bar S^\a] =\frac{1}{2}
\bar S^\b {{(\s_a)}_\b}^\a.
\label{algebra}
\eea
Here $(P,D,K,\mathcal{J}_a)$, $a=1,2,3$, are (Hermitian) bosonic generators of translations, dilatations, special conformal transformations, and $su(2)$ rotations, respectively. $Q_\alpha$ and $S_\alpha$ are fermionic generators of supersymmetry transformations and superconformal boosts, $\bar Q^\a$, $\bar S^\a$ being their Hermitian conjugates.
${{(\s_a)}_\b}^\a$ are the Pauli matrices (see Appendix).

Following the recipe in \cite{CWZ}, each generator in the superalgebra is then accompanied by a Goldstone superfield of the same Grassmann parity and the group--theoretic element is introduced
\bea
&&
\tilde g=e^{i t h} e^{\theta^\alpha q_\alpha+{\bar\theta}_\alpha {\bar q}^\alpha} e^{i\rho P} e^{\psi^\alpha Q_\alpha+{\bar\psi}_\alpha {\bar Q}^\alpha} e^{\phi^\alpha S_\alpha+{\bar\phi}_\alpha {\bar S}^\alpha} e^{i \mu  K} \times
\nonumber\\[2pt]
&&
\qquad
 e^{i \nu D} e^{i \lambda_a \mathcal{J}_a},
\label{elem}
\eea
in which $(\rho,\mu,\nu,\lambda_a)$ are real bosonic superfields and $(\psi_\alpha,\phi_\alpha,{\bar\psi}^\alpha,{\bar\phi}^\alpha)$ are complex fermionic superfields. Here $\rho$ and $\psi_\alpha$ are identified with those in the preceding section and the constraints (\ref{Const1}), (\ref{Const6}) are assumed to hold. The choice of $\tilde g$ is prompted by the study of the $d=1$, $\mathcal{N}=4$ superconformal mechanics in \cite{IKL}.

The left multiplication by a group element $\tilde g \to g\cdot \tilde g$
\be
g=e^{ia P} e^{\epsilon^\alpha Q_\alpha+{\bar\epsilon}_\alpha {\bar Q}^\alpha} e^{\kappa^\alpha S_\alpha+{\bar\kappa}_\alpha {\bar S}^\alpha} e^{i c  K} e^{i b D} e^{i \xi_a \mathcal{J}_a},
\ee
in which $(a,b,c,\xi_a)$ and $(\epsilon_\alpha,\kappa_\alpha)$ are bosonic and fermionic parameters, respectively, generates a finite $SU(1,1|2)$ transformation. In practical applications, it proves sufficient to focus on the infinitesimal transformations
\bea
&&
\rho'=\rho+a, \qquad \psi'_\alpha=\psi_\alpha, \qquad \bar\psi'^\alpha=\bar\psi^\alpha;
\nonumber\\[2pt]
&&
\rho'=\rho+b\rho, \quad \psi'_\alpha=\psi_\alpha+\frac 12 b \psi_\alpha, \quad \bar\psi'^\alpha=\bar\psi^\alpha+\frac 12 b \bar\psi^\alpha;
\nonumber\\[2pt]
&&
\rho'=\rho+c \rho^2-\frac 12 c \psi^2 \bar\psi^2, \quad \psi'_\alpha=\psi_\alpha+c \rho \psi_\alpha+\frac{i}{2} c \psi^2 \bar\psi_\alpha,
\nonumber\\[2pt]
&&
\bar\psi'^\alpha=\bar\psi^\alpha+c\rho \bar\psi^\alpha+\frac{i}{2} c \bar\psi^2 \psi^\alpha;
\nonumber\\[2pt]
&&
\rho'=\rho, \quad \psi'_\alpha=\psi_\alpha+\frac{i}{2} \xi_a {{(\s_a)}_\alpha}^\beta \psi_\beta,
\nonumber\\[2pt]
&&
\bar\psi'^\alpha=\bar\psi^\alpha-\frac{i}{2} \xi_a \bar\psi^\beta {{(\s_a)}_\beta}^\alpha  ;
\nonumber\\[2pt]
&&
\rho'=\rho+i\left(\bar\psi\epsilon-\bar\epsilon\psi\right), \quad \psi'_\alpha=\psi_\alpha+\epsilon_\alpha, \quad \bar\psi'^\alpha=\bar\psi^\alpha+\bar\epsilon^\alpha;
\nonumber\\[2pt]
&&
\rho'=\rho-i\rho \left(\bar\psi \kappa-\bar\kappa \psi \right)+\bar\psi \psi \left(\bar\psi \kappa+\bar\kappa \psi \right),
\nonumber\\[2pt]
&&
\psi'_\alpha=\psi_\alpha-\rho \kappa_\alpha+i\bar\psi\psi \kappa_\alpha-i\psi^2 \bar\kappa_\alpha,
\nonumber\\[2pt]
&&
\bar\psi'^\alpha=\bar\psi^\alpha-\rho \bar\kappa^\alpha-i\bar\psi\psi \bar\kappa^\alpha-i\bar\psi^2 \kappa^\alpha,
\label{tr}
\eea
which are obtained with the aid of the Baker--Campbell--Hausdorff formula
\be
e^{iA}~ T~ e^{-iA}=T+\sum_{n=1}^\infty\frac{i^n}{n!}
\underbrace{[A,[A, \dots [A,T] \dots]]}_{n~\rm times}.
\label{ser}
\ee
Note that both the original and transformed superfields depend on the same arguments $(t,\theta,\bar\theta)$ such that the transformations affect the form of the superfields only, e.g. $\delta \rho=\rho'(t,\theta,\bar\theta)-\rho(t,\theta,\bar\theta)$.
Computing the algebra of the infinitesimal transformations (\ref{tr}), one can verify that it does reproduce the structure relations (\ref{algebra}).

As the next step, one builds the odd analogues of the Maurer--Cartan one--forms
\bea
&&
{\tilde g}^{-1} \mathcal{D}^\alpha \tilde g=i \omega_D^\alpha D+i \omega_K^\alpha K+\omega_Q^{\alpha\beta} Q_\beta+\omega_S^{\alpha\beta} S_\beta+
\nonumber\\[2pt]
&&
\qquad \qquad \quad
{{\omega_{\bar S}}_\beta}^\alpha {\bar S}^\beta+i \omega_{\mathcal{J}}^{\alpha a} \mathcal{J}_a-q^\alpha,
\eea
where $\mathcal{D}^\alpha$ is the covariant derivative (\ref{covd}), which give rise to the $SU(1,1|2)$ invariants
\bea
&&
\omega_D^\alpha=\mathcal{D}^\alpha \nu+2i \bar\phi^\beta \mathcal{D}^\alpha \psi_\beta,
\nonumber\\[8pt]
&&
\omega_K^\alpha=e^\nu \left(\mathcal{D}^\alpha \mu-2i \mu \bar\phi^\beta \mathcal{D}^\alpha \psi_\beta -i\bar\phi^\beta \mathcal{D}^\alpha \phi_\beta  -
\right.
\nonumber\\[2pt]
&&
\qquad \quad
\left.
\bar\phi^2 \phi^\beta \mathcal{D}^\alpha \psi_\beta-i\phi_\beta \mathcal{D}^\alpha \bar\phi^\beta \right),
\nonumber\\[8pt]
&&
\omega_Q^{\alpha\gamma}=e^{-\frac{\nu}{2}} \mathcal{D}^\alpha \psi^\beta {{\left(\mbox{exp} \left[\frac{i}{2} \lambda_c \sigma_c\right] \right)}_\beta}^\gamma,
\nonumber\\[2pt]
&&
\omega_S^{\alpha\gamma}=e^{\frac{\nu}{2}} \left(\mathcal{D}^\alpha \phi^\beta+\mu \mathcal{D}^\alpha \psi^\beta -i \bar\phi \phi \mathcal{D}^\alpha \psi^\beta \right) \times
\nonumber\\[2pt]
&&
\qquad \quad
{{\left(\mbox{exp} \left[\frac{i}{2} \lambda_c \sigma_c\right] \right)}_\beta}^\gamma,
\nonumber\\[2pt]
&&
{{\omega_{\bar S}}_\gamma}^\alpha=e^{\frac{\nu}{2}} \left(\mathcal{D}^\alpha \bar\phi_\beta+i\bar\phi^2 \mathcal{D}^\alpha \psi_\beta \right) \times
\nonumber\\[2pt]
&&
\qquad \quad
{{\left(\mbox{exp} \left[-\frac{i}{2} \lambda_c \sigma_c\right] \right)}_\gamma}^\beta,
\label{invariants}
\eea
while $i \omega_{\mathcal{J}}^{\alpha a} \mathcal{J}_a$ yields
\bea
e^{i\lambda_c \mathcal{J}_c} \left(\mathcal{D}^\alpha e^{-i\lambda_k \mathcal{J}_k} \right)+2i \left( \mathcal{D}^\alpha \psi^\beta {{(\s_a)}_\beta}^\gamma \bar\phi_\gamma \right) \mathcal{J}_a.
\label{om}
\eea
The invariant ${\tilde g}^{-1} \bar{\mathcal{D}}_\alpha \tilde g$ could be considered likewise, which would result in the Hermitian conjugates of (\ref{invariants}) and (\ref{om}).

Because the $\mathcal{N}=4$ super-Schwarzian derivative is expected to involve the fermionic superfield $\psi_\alpha$ only, one is led to use
the invariants (\ref{invariants}) so as to eliminate $(\nu,\mu,\lambda_a,\phi_\alpha,\bar\phi^\alpha)$ from the consideration.
Guided by a recent analysis of the $\mathcal{N}=1$ and $\mathcal{N}=2$ super-Schwarzian derivatives in a similar setting \cite{AG}, let us impose the following constraints
\be
\omega_D^\alpha=0, \quad \omega_Q^{\alpha\gamma}=r^{\alpha\gamma}, \quad \omega_S^{\alpha\gamma}=0,
\label{const3}
\ee
where $r^{\alpha\gamma}$ is a constant matrix with even supernumber elements (coupling constants). Note that the consistency requires $r^{\alpha\beta}$ and its Hermitian conjugate ${\left(r^{\alpha\beta}\right)}^{\dagger}={\bar r}_{\beta\alpha}$ to obey the restrictions ${\bar r}_{\alpha\beta} r^{\beta\gamma}=\frac 12 {\delta_\alpha}^\gamma (r {\bar r})$, $r^{\alpha\beta} {\bar r}_{\beta\gamma} =\frac 12 {\delta_\gamma}^\alpha (r {\bar r})$ (see Eq. (\ref{La}) below).

Making use of the Hermitian conjugates, one can express $\nu$ and $\mu$ in terms of the fermionic superfields $\psi_\alpha$ and $\phi_\alpha$
\bea
&&
e^\nu=\frac{\left(\mathcal{D} \psi \bar{\mathcal{D}} \bar\psi \right)}{r \bar r},
\nonumber\\[2pt]
&&
\mu \left( \mathcal{D} \psi \bar{\mathcal{D}} \bar\psi\right)+\left(\mathcal{D} \phi \bar{\mathcal{D}} \bar\psi\right)-i\bar\phi \phi \left(\mathcal{D} \psi \bar{\mathcal{D}} \bar\psi\right)=0,
\label{SOL}
\eea
where $r {\bar r}=r^{\alpha\beta} {\bar r}_{\beta \alpha}$, while $\omega_Q^{\alpha\gamma}$ links $\lambda_a$ to $\psi_\alpha$
\be
{{\left(\mbox{exp} \left[\frac{i}{2} \lambda_a \sigma_a\right] \right)}_\alpha}^\beta=-\frac{2 e^{\frac{\nu}{2}} r^{\gamma\beta}  \bar{\mathcal{D}}_\gamma \bar\psi_\alpha }{\mathcal{D} \psi \bar{\mathcal{D}} \bar\psi}.
\label{La}
\ee

In order to link $(\phi_\alpha,{\bar\phi}^\alpha)$ to $(\psi_\alpha,{\bar\psi}^\alpha)$, it suffices to contract $\omega_D^\alpha=0$ with $\bar{\mathcal{D}}_\alpha \bar\psi^\gamma$ and to make use of the quadratic constraints (\ref{Const5}). The result reads
\be
\phi_\alpha=-\frac{2\partial_t \psi_\alpha}{\left(\mathcal{D} \psi \bar{\mathcal{D}} \bar\psi\right)}, \qquad  \bar\phi^\alpha=-\frac{2\partial_t \bar\psi^\alpha}{\left(\mathcal{D} \psi \bar{\mathcal{D}} \bar\psi\right)},
\label{phi}
\ee
which also simplifies the expression for $\mu$
\be
\mu=
2 \frac{\left(\mathcal{D}^\alpha \partial_t  \psi_\beta \right) \left(\bar{\mathcal{D}}_\alpha \bar\psi^\beta\right)}{{\left(\mathcal{D} \psi \bar{\mathcal{D}} \bar\psi \right)}^2}.
\label{mu}
\ee

Finally, substituting (\ref{SOL}), (\ref{La}), (\ref{phi}), (\ref{mu}) back into the invariants (\ref{invariants}), one reveals that they either vanish identically
(it proves helpful to evaluate $\omega_{\mathcal{J}}^{\alpha a} \mathcal{J}_a$  at $\mathcal{J}_a=\frac 12 \s_a$), or produce the second--rank tensor
\bea
&&
\bar{\mathcal{D}}_\alpha \bar\psi^\gamma \mathcal{D}^\beta \left(\frac{\partial_t \psi_\gamma}{\mathcal{D} \psi \bar{\mathcal{D}} \bar\psi} \right)
-\frac 12 {\delta_{\alpha}}^\beta \left( \bar{\mathcal{D}}_\mu \bar\psi^\nu \mathcal{D}^\mu\left(\frac{\partial_t \psi_\nu}{\mathcal{D} \psi \bar{\mathcal{D}} \bar\psi} \right) \right):=
\nonumber\\[2pt]
&&
{\mathcal{I}_\alpha}^\beta,
\label{I}
\eea
or, lowering the upper index,
\be
\mathcal{I}_{\alpha\beta}=\frac{1}{8i} \left(\mathcal{D}_\alpha \bar{\mathcal{D}}_\beta+\mathcal{D}_\beta \bar{\mathcal{D}}_\alpha \right) \ln{\left(\mathcal{D} \psi \bar{\mathcal{D}} \bar\psi\right)}.
\label{III}
\ee
Thus, after the auxiliary fields $(\mu,\nu,\lambda_a,\phi_\alpha,{\bar\phi}^\alpha)$ have been eliminated form the consideration, the only $SU(1,1|2)$ invariant combination involving $\psi_\alpha$ is given by (\ref{III}). Because ${\mathcal{I}_\alpha}^\beta$ is traceless, its contains three independent components which can be extracted by contracting with the (traceless) Pauli matrices
\be
\mathcal{D}^\alpha {{(\s_a)}_\a}^\b \bar{\mathcal{D}}_\beta \ln{\left(\mathcal{D} \psi \bar{\mathcal{D}} \bar\psi\right)}:=\mathcal{I}_a[\psi(t,\theta,\bar\theta);t,\theta,\bar\theta].
\label{N4S}
\ee
Eq. (\ref{N4S}) coincides with the $\mathcal{N}=4$ super--Schwarzian derivative introduced in \cite{MU}. Note that it transforms as a vector under $SU(2)$ transformations acting in $\mathcal{S}^{1|4}$ superspace. It also satisfies the conditions
\be
\mathcal{D}^2 \mathcal{I}_a=\bar{\mathcal{D}}^2 \mathcal{I}_a=0.
\ee

A few comments are in order. Firstly,
focusing on the infinitesimal $SU(1,1|2)$ transformations (\ref{tr}) and taking into account the constraints (\ref{Const1}) and the equalities
\bea
&&
\mathcal{D} \psi' \bar{\mathcal{D}} \bar\psi'=(1+2 c \rho) \mathcal{D} \psi \bar{\mathcal{D}} \bar\psi, \quad \mathcal{D} \psi' \bar{\mathcal{D}} \bar\psi'=(1+b) \mathcal{D} \psi \bar{\mathcal{D}} \bar\psi,
\nonumber\\[2pt]
&&
\mathcal{D} \psi' \bar{\mathcal{D}} \bar\psi'=\left(1+2i(\bar\kappa\psi-\bar\psi\kappa)\right) \mathcal{D} \psi \bar{\mathcal{D}} \bar\psi,
\eea
one can verify that (\ref{N4S}) does hold invariant.

Secondly,
considering the superconformal diffeomorphism described by
Eqs. (\ref{sdiff}), (\ref{Const1}), and (\ref{Const3}) above, changing the argument $\psi_\alpha(t,\theta,\bar\theta) \to \Omega_\alpha (\rho,\psi,\bar\psi)$ of the super--Schwarzian (\ref{N4S}), and taking into account Eq. (\ref{hom1}), which gives rise to the identity
\be
\left(\mathcal{D} \Omega \bar{\mathcal{D}} \bar\Omega\right)=\frac 12 \left(\mathcal{D} \psi \bar{\mathcal{D}} \bar\psi\right) \left(\mathcal{D}' \Omega \bar{\mathcal{D}}' \bar\Omega\right),
\ee
one reveals the composition law \cite{MU}
\bea
&&
\mathcal{I}_a[\Omega(t',\theta',\bar\theta');t,\theta,\bar\theta]=\mathcal{I}_a[\psi(t,\theta,\bar\theta);t,\theta,\bar\theta]+
\nonumber\\[2pt]
&&
\qquad \qquad \qquad \quad
\frac 12 M_{ab}
\mathcal{I}_b[\Omega(t',\theta',\bar\theta');t',\theta',\bar\theta'],
\label{TRL}
\eea
with $M_{ab}=\mathcal{D}^\alpha \psi_\lambda  {{(\s_a)}_\a}^\b  {{(\s_b)}_\nu}^\lambda  \bar{\mathcal{D}}_\beta \bar\psi^\nu$. In deriving Eq. (\ref{TRL}), the properties of the Pauli matrices exposed in Appendix have been used.
In particular, the $\mathcal{N}=4$ super--Schwarzian holds invariant under the change of the argument $\psi_\alpha(t,\theta,\bar\theta) \to \Omega_\alpha (\rho,\psi,\bar\psi)$, provided the last term in (\ref{TRL}) vanishes.

Thirdly, assuming the constraints (\ref{Const1}) and (\ref{Const6}) to hold, which result in the restrictions (\ref{funct})
upon the components of the chiral superfield (\ref{csf}) (recall that $v$ in (\ref{funct}) was set to zero),
and analysing the equation $\mathcal{I}_a[\psi(t,\theta,\bar\theta);t,\theta,\bar\theta]=0$, or equivalently $\mathcal{I}_{\alpha\beta}=0$,
one can fix $u(t)$ and $\alpha_\gamma (t)$
\be
u(t)=\frac{1}{c t +d}, \qquad \alpha_\gamma(t)=\epsilon_\gamma+\frac{(\bar\kappa \kappa) \kappa_\gamma}{ct+d},
\ee
where $(c,d)$ and $(\epsilon_\gamma,\kappa_\gamma)$ are bosonic and fermionic parameters, respectively. The resulting superfield (\ref{csf}) determines a finite $SU(1,1|2)$ transformation acting in the odd sector of $\mathcal{S}^{1|4}$ superspace. In particular, it correctly reduces to Eqs. (\ref{tr}) in the infinitesimal limit \cite{footnote2}. A finite $SU(1,1|2)$ transformation acting in the even sector of $\mathcal{S}^{1|4}$ can be found by integrating Eq. (\ref{cond1}).

\section{\label{sec:IV} Conclusion}

To summarize, in this work we applied the method of nonlinear realizations to the superconformal group $SU(1,1|2)$ and reproduced the $\mathcal{N}=4$ super--Schwarzian derivative in \cite{MU}. As compared to other approaches in the literature, our consideration relied upon invariants of $SU(1,1|2)$ alone. Nether infinite--dimensional extension of the supergroup, nor conformal field theory techniques, nor the analysis of central charges/cocycles were needed.

Turning to possible further developments, it would be interesting to study a variant of an $\mathcal{N}=4$ supersymmetric extension of the Sachdev--Ye--Kitaev model which is based upon the effective action
involving $\mathcal{I}_a \mathcal{\bar I}_a$.

Apart from the $\mathcal{N}=4$ super--Schwarzian associated with $SU(1,1|2)$, one can define a similar derivative which exhibits $OSp(4|2)$ superconformal invariance \cite{Sch}. Both the supergroups are known to be particular instances of the most general superconformal group in one dimension $D(2,1;\alpha)$. To the best of our knowledge, a $D(2,1;\alpha)$ super--Schwarzian has not yet been constructed and we hope to report on its peculiarities elsewhere.

Another interesting problem is to obtain an $\mathcal{N}=3$ super--Schwarzian along similar lines.

\begin{acknowledgments}

This work is supported by the Russian Foundation for Basic Research, grant No 20-52-12003.

\end{acknowledgments}

\vspace{0.5cm}

\appendix

\section{$SU(2)$ spinor conventions}

Throughout the text we use a lower Greek index to designate an $SU(2)$--doublet representation. Hermitian conjugation
yields an equivalent representation to which one assigns an upper index
\be
{(\p_\a)}^{\dagger}={\bar\p}^\a\ , \qquad \a=1,2\ .
\nonumber
\ee
As usual, spinor indices are raised and lowered with the use of the $SU(2)$--invariant
antisymmetric matrices
\be
\p^\a=\e^{\a\b}\p_\b\ , \quad {\bar\p}_\a=\e_{\a\b} {\bar\p}^\b\ ,
\nonumber
\ee
where $\e_{12}=1$, $\e^{12}=-1$. For spinor bilinears we stick to the notation
\be
\quad \p^2=(\p^\a \p_\a\ ) , \quad
\bar\p^2=(\bar\p_\a \bar\p^\a )\ , \quad \bar\p \p=(\bar\p^\a \p_\a )\ ,
\nonumber
\ee
such that
\bea
&&
\p_\a \p_\b=\frac 12 \e_{\a\b} \p^2\ , \qquad \bar\p^\a \bar\p^\b=\frac 12 \e^{\a\b} \bar\p^2\ ,
\nonumber\\[2pt]
&&
\p_\a \bar\p_\b-\p_\b \bar\p_\a=\e_{\a\b}
(\bar\p \p)\ , \qquad \psi^\alpha \psi^\beta=-\frac 12 \epsilon^{\alpha\beta} \psi^2,
\nonumber\\[2pt]
&&
\bar\psi_\alpha \bar\psi_\beta=-\frac 12 \epsilon_{\alpha\beta} \bar\psi^2, \quad {\left(\bar\psi \psi \right)}^2=\frac 12 \psi^2 \bar\psi^2.
\nonumber
\eea

The Pauli matrices ${{(\s_a)}_\a}^\b$
are taken in the standard form
\be
\s_1=\begin{pmatrix}0 & 1\\
1 & 0
\end{pmatrix}\ , \qquad \s_2=\begin{pmatrix}0 & -i\\
i & 0
\end{pmatrix}\ ,\qquad
\s_3=\begin{pmatrix}1 & 0\\
0 & -1
\end{pmatrix}\ ,
\nonumber
\ee
which obey
\bea
&&
{{(\s_a \s_b)}_\a}^\b +{{(\s_b \s_a)}_\a}^\b=2 \d_{ab} {\d_\a}^\b \ ,
\nonumber\\[2pt]
&&
{{(\s_a \s_b)}_\a}^\b -{{(\s_b \s_a)}_\a}^\b=2i \e_{abc} {{(\s_c)}_\a}^\b \ ,
\nonumber\\[2pt]
&&
{{(\s_a \s_b)}_\a}^\b=\d_{ab} {\d_\a}^\b +i \e_{abc} {{(\s_c)}_\a}^\b \ ,
\nonumber\\[2pt]
&&
{{(\s_a)}_\a}^\b {{(\s_a)}_\g}^\r=2 {\d_\a}^\r {\d_\g}^\b-{\d_\a}^\b {\d_\g}^\r\ ,
\nonumber\\[2pt]
&&
{{(\s_a)}_\a}^\b \e_{\b\g} ={{(\s_a)}_\g}^\b \e_{\b\a}\ ,
\nonumber\\[2pt]
&&
 \e^{\a\b} {{(\s_a)}_\b}^\g=\e^{\g\b} {{(\s_a)}_\b}^\a \ ,
 \nonumber
\eea
where $\e_{abc}$ is the totally antisymmetric Levi-Civit\'a tensor, $\e_{123}=1$. From the last line one finds
\bea
&&
{{\left(\mbox{exp} \left[-\frac{i}{2} \xi_a \sigma_a \right] \right)}_\alpha}^\gamma \epsilon_{\gamma\beta}=-{{\left(\mbox{exp} \left[\frac{i}{2} \xi_a \sigma_a\right] \right)}_\beta}^\gamma \epsilon_{\gamma\alpha},
\nonumber\\[2pt]
&&
\epsilon^{\alpha\gamma} {{\left(\mbox{exp} \left[-\frac{i}{2} \xi_a \sigma_a \right] \right)}_\gamma}^\beta =-\epsilon^{\beta\gamma} {{\left(\mbox{exp} \left[\frac{i}{2} \xi_a \sigma_a\right] \right)}_\gamma}^\alpha,
\nonumber
\eea
where $\xi_a$ is a real vector parameter.

Throughout the text we use the abbriviation
$\bar\p \s_a \p=\bar\p^\a {{(\s_a)}_\a}^\b \p_\b$.
Our convention for the Hermitian conjugation adopted above imply
\bea
&&
{(\bar\psi_\alpha)}^{\dagger}=-\psi^\alpha\ , \qquad
{(\psi^2)}^{\dagger}=\bar\psi^2\ , \qquad {(\bar\psi\,\sigma_a \chi)}^{\dagger}=\bar\chi \sigma_a \psi\ .
\nonumber
\eea

\begin{thebibliography}{nn}
\bibitem{FGMS}
W. Fu, D. Gaiotto, J. Maldacena, S. Sachdev, {\it Supersymmetric
Sachdev--Ye--Kitaev models}, Phys. Rev. D {\bf 95} (2017) 026009,  arXiv:1610.08917.
\bibitem{MTV}
T.G. Mertens, G.J. Turiaci, H.L. Verlinde, {\it Solving the Schwarzian via the conformal bootstrap}, JHEP {\bf 1708} (2017) 136, arXiv:1705.08408.
\bibitem{BBN}
M. Berkooz, N. Brukner, V. Narovlansky, A. Raz, {\it The double scaled limit of super--symmetric SYK models}, arXiv:2003.04405.
\bibitem{F}
D. Friedan, {\it Notes on string theory and two--dimensional conformal field theory}, Unified String Theories: proceedings (ed. by M.B. Green and D.J. Gross). Singapore, World Scientific, 1985.
\bibitem{Cohn}
J.D. Cohn, {\it $\mathcal{N}=2$ super--Riemann surfaces}, Nucl. Phys. B {\bf 284} (1987) 349.
\bibitem{Sch}
K. Schoutens, {\it $O(n)$ extended superconformal field theory in superspace}, Nucl. Phys. B {\bf 295} (1988) 634.
\bibitem{MU}
S. Matsuda,  T. Uematsu, {\it Super  Schwarzian  derivatives  in $\mathcal{N}=4$ $su(2)$-extended  superconformal algebras}, Mod. Phys. Lett. A {\bf 11} (1990) 841.
\bibitem{footnote}
Because conformal transformations in $\mathcal{R}^1$ involve the inversion $t \to \frac{1}{t}$,
the conformal group $SL(2,R)$ does not act globally on $\mathcal{R}^1$, but rather on $\mathcal{S}^1=\mathcal{RP}^1$.
\bibitem{MD}
J.P. Michel, C. Duval, {\it On the projective geometry of the supercircle: a unified construction of the super cross--ratio and Schwarzian derivative}, Int. Math. Res. Not. 2008 (2008) 054, arXiv:0710.1544.
\bibitem{ChK}
D. Chang, A. Kumar, {\it Representations of $\mathcal{N}=3$
superconformal algebra}, Phys. Lett. B {\bf 193} (1987) 181.
\bibitem{AG}
A. Galajinsky, {\it Super-Schwarzians via nonlinear realizations}, JHEP {\bf 2006} (2020) 027, arXiv:2004.04489.
\bibitem{CWZ}
S.R. Coleman, J. Wess, B. Zumino, {\it Structure of phenomenological Lagrangians. I}, Phys. Rev. {\bf 177} (1969) 2239.\\
C.G. Callan, S.R. Coleman, J. Wess, B. Zumino, {\it Structure of phenomenological Lagrangians. II}, Phys. Rev. {\bf 177} (1969) 2247.\\
D. Volkov, {\it Phenomenological lagrangians}, Sov. J. Part. Nucl. {\bf 4} (1973) 3.
\bibitem{IO}
E.A. Ivanov, V.I. Ogievetsky, {\it The inverse Higgs phenomenon in
nonlinear realizations}, Theor. Math. Phys. {\bf 25} (1975) 1050.
\bibitem{AG1}
A. Galajinsky, {\it Schwarzian mechanics via nonlinear realizations}, Phys. Lett. B {\bf 795} (2019) 277, arXiv:1905.01935.
\bibitem{IKL1}
E. Ivanov, S. Krivonos, V. Leviant, {\it  Geometric superfield approach to superconformal mechanics},
J. Phys. A {\bf 22} (1989) 4201.
\bibitem{IKL}
E. Ivanov, S. Krivonos, O. Lechtenfeld, {\it New variant of $\mathcal{N}=4$ superconformal mechanics}, JHEP {\bf 03} (2003) 014, hep-th/0212303.
\bibitem{footnote2}
In order to reproduce the infinitesimal form of the superconformal boosts entering (\ref{tr}), one sets $d=1$, considers $c$ to be small, such that $\frac{1}{1+c t}\approx  1-c t$, and identifies $c (\bar\kappa \kappa) \kappa_\gamma$ with the infinitesimal $\kappa_\gamma$ in (\ref{tr}). The resulting transformation is a superposition of the supersymmetry transformation, special conformal transformation parametrized by $c$ and the superconformal boost associated with $c (\bar\kappa \kappa) \kappa_\gamma$.

\end{thebibliography}

\end{document}